\documentclass[11pt]{article}
\usepackage[margin=1in]{geometry}
\usepackage{makecell}
\usepackage[T1]{fontenc}
\usepackage[utf8]{inputenc}
\usepackage{booktabs}
\usepackage{amsmath}
\usepackage{amsfonts}
\usepackage[hyphens]{url}
\usepackage{hyperref}
\hypersetup{
    colorlinks=true,
    linkcolor=blue,
    filecolor=magenta,      
    urlcolor=cyan,
    citecolor=blue,
    pdftitle={Symmetry in Software Platforms as an Architectural Principle},
    pdfauthor={Bjørn Remseth},
    pdfsubject={Software Engineering, Distributed Systems, Platform Architecture},
    pdfkeywords={software platforms, symmetry, architectural principles, Kubernetes, platform design, software architecture, distributed systems, container orchestration, system invariants, platform evolution},
    bookmarksopen=true,
    bookmarksnumbered=true
}

\newcommand{\builddate}{2025-10-22}

\newcommand{\versionstring}{Version: 2025-10-22 (6d08b5d)}

\begin{document}

\title{Symmetry in Software Platforms as an Architectural Principle}
\author{Bj\o rn Remseth\\Microsoft\\bjornremseth@microsoft.com}

\newcommand{\arxivsubjects}{
\textbf{Subject classifications:} cs.SE (Software Engineering)
}

\newcommand{\documentversion}{\ifdefined\builddate\builddate\else\today\fi}
\newcommand{\workingpaper}{
\textbf{Working Paper Notice:} This is a working paper submitted to arXiv for community feedback and discussion. Comments and suggestions are welcome.
}

\newcommand{\affiliationdisclaimer}{ \textbf{Affiliation Disclaimer:}
  While the author is affiliated with Microsoft, this work is
  conducted independently and is not connected to any Microsoft
  commercial activities. This paper represents the author's personal
  research interests.  }

\date{\documentversion}

\maketitle

\noindent \arxivsubjects

\vspace{0.5em}
\noindent \workingpaper

\vspace{0.5em}
\noindent \affiliationdisclaimer

\ifdefined\versionstring
\vspace{0.5em}
\noindent \textit{\small \versionstring}
\fi

\begin{abstract}

Software platforms often act as \emph{structure-preserving
systems}. They provide consistent interfaces and behaviors that remain
stable under specific transformations.  This paper explores the idea
that architectural robustness emerges from enforcing such structural
regularities.
  
We begin by reviewing earlier work on invariance and uniformity in
software architecture and platform design.  Next, we formalize these
ideas using analogies from group theory, focusing on transformation
invariants such as component interchangeability, temporal stability,
and interface uniformity.  A case study of Kubernetes illustrates how
these principles are realized in practice.

Kubernetes maintains several forms of invariance. It treats differend
kinds of nodes and pods as equivalent elements and exposes a uniform
API surface.  At the same time, it introduces selective departures
from uniformity through roles, affinities, and scheduling
constraints where differentiation adds value.

We conclude by examining how such structural consistency and
controlled variation influence platform evolution and governance.
Preserving regularity simplifies design and long-term maintenance,
while deliberate exceptions accommodate specialized needs.  Finally,
we outline future directions for research on how architectural
invariants can be leveraged to strengthen platform design.

\end{abstract}
  

\section{Introduction}
\label{sec:introduction}

Software platforms ranging from operating systems and cloud
frameworks to container orchestration systems provide a stable
foundation upon which applications and services can be built. A key
challenge in platform design is balancing uniformity and flexibility:
the platform must present a consistent set of behaviors and interfaces
to many different clients and components, yet also accommodate change,
scaling, and heterogeneity. In this paper, we explore the view that
this balance is achieved by treating software platforms as
\emph{symmetry-enforcing systems}, which impose invariants or
equivalences that hold even as various aspects of the system
change. The term \emph{symmetry} in this context refers to
transformations or changes that can be applied to the system without
altering certain essential properties or outcomes.  In this paper, we
define a symmetry-enforcing system as a software platform that either
deliberately imposes or naturally exhibits invariants under specific
transformations. This notion can be used both as a design methodology
(when consciously applied) and as an analytical framework (when
retrospectively identifying symmetries in existing systems).

This idea is inspired by the use of symmetry in mathematics and
physics, where a system is symmetric if some transformation
(e.g. rotation, reflection, time-shift) leaves the system's observable
behavior unchanged. In software, symmetry manifests as the ability to
substitute one component for another, repeat an interaction at a
different time or location, or scale the system up or down, all
without affecting the platform's correctness or fundamental
behavior. Prior work suggests that software architectures benefit
from symmetry through greater stability and flexibility in the face of
change. For example, Arsanjani \cite{Arsanjani2001} defines symmetry
in software architecture as \emph{``invariance to change´´} – the
capacity of an architecture to remain stable along certain axes of
variation. Zhao and Coplien \cite{Zhao2003} similarly describe
symmetry as \emph{``the possibility of making a change while some
aspect remains immune to this change.''}

The hypothesis guiding this work is that successful platforms enforce
symmetry in their core design. By identifying and codifying
assumptions that remain invariant, for instance, that any server node
can host any service, or that any client will see the same interface.
Platforms achieve a form of regularity that simplifies reasoning and
integration. At the same time, absolute symmetry is neither possible
nor always desirable, platforms often must \emph{break} symmetry in
controlled ways to introduce specializations or optimizations
analogous to symmetry breaking in physics leading to new
structure.

\begin{table}[h]
  \small
  \caption{Two perspectives on symmetry-enforcing systems: as a design methodology and as an analytical framework.}
  \label{tab:symmetry-perspectives}
  \begin{tabular}{p{2.0cm}p{5.5cm}p{5.5cm}}
    \toprule
    \textbf{Dimension} & \textbf{Design Methodology} & \textbf{Analytical Framework} \\
    \midrule
    Goal & Build systems with desirable invariants & Identify and reason about existing invariants \\
    
    Use case & Architecture planning, API design & System evolution, refactoring, auditing \\
    
    Symmetry role & Target to enforce & Observation for insight or refactor \\
    
    Examples & Use of uniform APIs, CRD patterns & Detecting unintentional special-case behavior \\
    \bottomrule
  \end{tabular}
\end{table}

We use the term symmetry-enforcing system to describe a software
platform that either deliberately imposes invariants or exhibits
natural regularities that persist under transformation. This concept
can be understood in two complementary ways: as a \emph{design
methodology}, where symmetry principles guide the proactive
construction of platforms to support modularity, interchangeability,
and robustness; and as an \emph{analytical framework}, where symmetry
is used to interpret and decompose existing systems by identifying
invariants and understanding where and why symmetry breaks. By
treating symmetry both as a prescriptive tool and a descriptive lens,
we aim to articulate not only how platforms can be built to leverage
symmetry, but also how existing systems can be analyzed for structural
coherence and evolutionary constraints. In table
\ref{tab:symmetry-across-domains} we illustrate different types of
symmetries in Kubernetes, ERP systems, and social media platforms.

This paper aims to: (1) review how symmetry has been treated in
software architecture literature, (2) formalize a notion of symmetry
in platform design, (3) apply this formalism to analyze Kubernetes, a
widely-used cloud platform, and (4) discuss the broader implications
of symmetry and symmetry-breaking for platform evolution and
governance. The remainder of this paper is organized as
follows. Section \ref{sec:literatureReview} reviews related
work. Section \ref{sec:formalizingSymmetryInSoftwarePlatforms}
formalizes the concept of symmetry. Section
\ref{sec:caseStudyKubernetes} presents the Kubernetes case
study. Section \ref{sec:discussion} discusses the implications for
platform design, and section \ref{sec:conclusionsAndFutureWork}
concludes and suggests future work.

Although Kubernetes serves as our primary case study, the principle of
symmetry applies more broadly. From the uniform transaction models in
enterprise systems, to symmetric dataflow and event models in social
media platforms, to the structural invariants in programming language
runtimes, symmetry recurs as a design-enabling constraint. We focused
on Kubernetes to ground our discussion, but the broader relevance of
symmetry motivates future comparative studies across diverse platform
domains.

\section{Literature Review}
\label{sec:literatureReview}

\subsection{Symmetry in Software Architecture and Design}
The concept of symmetry in software has been explored by multiple
researchers as a means to understand and improve software
structure. As mentioned earlier, Arsanjani\cite{Arsanjani2001}
introduced the notion of \emph{software symmetry} in n-tier
distributed architectures. In this view, a software system exhibits
symmetry if certain structural or behavioral properties remain
unchanged when the system undergoes transformations such as
reconfiguration, scaling, or component replacement. A symmetric
architecture can better withstand changes in requirements or usage
volume without requiring fundamental redesign. Invariant “axes of
symmetry” might include the ability to add more users (scalability) or
to swap out data sources, all while preserving system stability. This
work drew an analogy to physics: just as breaking perfect symmetry
yields complex structures in nature, breaking symmetry in software can
introduce instability if not done in a balanced way.

Zhao and Coplien’s work in the early 2000s further formalized symmetry
in object-oriented software design. In \emph{Understanding Symmetry in
Object-Oriented Languages}\cite{Zhao2003}, they define symmetry as
“the possibility of making a change while leaving other key aspects
invariant.” They illustrate this with the concept of classes: a class
defines a set of objects that are interchangeable in terms of their
interface and behavior. Any instance can be substituted for any other
(symmetry under permutation of instances) so long as the class’s
contract is respected. This symmetry aids correctness and
optimization for example, a compiler can generate a single code layout
for all instances. This interchangeability is closely related to the
Liskov Substitution Principle\cite{liskov1994behavioral} and subtype
polymorphism.

Symmetry has been discussed in the context of design patterns:
Alexander’s pattern language in
architecture, and its influence on software
patterns\cite{beck1987pattern}) implicitly values symmetry-breaking
and symmetry-forming processes. Liping Zhao\cite{Zhao2008} argued that
many software design patterns can be seen as outcomes of
\emph{symmetry breaking}. Programming languages provide symmetric,
uniform abstractions (all objects of a class being structurally
identical). Patterns often introduce purposeful irregularities on top
of these uniform abstractions. For example, the Singleton pattern
breaks the symmetry of class instantiation by ensuring only one
instance exists. This reduction in symmetry is controlled: it solves a
specific problem while retaining other invariants.

\subsection{Symmetry in Platforms and Systems Theory}
Beyond object-oriented design, symmetry appears in broader systems and
platform theory. Distributed systems often leverage symmetry by
treating nodes or processes uniformly. In a peer-to-peer network or a
distributed algorithm like Paxos\cite{lamport1998paxos}, all
participants may run the same code and play interchangeable roles,
achieving fault tolerance through symmetric redundancy. Only when
necessary is symmetry broken (e.g., electing a leader), a temporary
asymmetry required to avoid deadlock or indecision.

In software \emph{platforms}, symmetry can manifest as standardized
interfaces and protocols. A classic example is the REST architecture:
Fielding’s \emph{uniform interface} constraint mandates that the same
set of methods (e.g., HTTP GET, POST, PUT, DELETE) are used in a
consistent way for all resources. This is a symmetry principle: every
resource is accessed through the same invariant set of operations,
decoupling client and server implementations and simplifying the
platform's overall complexity. Similarly, in operating systems, the
“everything is a file” paradigm in Unix is a symmetry that provides a
uniform way to read/write diverse resources, leading to simple
compositions and generic tooling.

Platform stability and evolvability often come from identifying the
right abstractions that remain stable. As Mihai\cite{mihai2019}
observed, successful system designers “identify the possible
symmetries and invariances in the system” to simplify the
problem. This resonates with the idea of \emph{conserved quantities}
in physics: Noether’s theorem famously states that every symmetry of a
Lagrangeian describing a physical system corresponds to a conserved
quantity.  We draw inspiration from Noether’s Theorem. Software
lacks a formal Lagrangian structure, but the analogy encourages us to ask:
what invariants are preserved under change?  In software platforms,
one might say that invariants hint at underlying symmetries. For
instance, the backward compatibility of an API can be viewed as a time
symmetry: the platform’s behavior is invariant with respect to
time. The “conserved quantity” is the client’s ability to function.

With the exception of quantum computer systems\cite{Benioff1980,
  Nielsen2010} which are outside of our scope, software systems are
not commonly described by Lagrangians. We therefore apply reasoning by
analogy rather than a direct invocation of Noether's Theorem with all
its mathematical preconditions. In the context of software platforms,
invariants can be seen as hinting at underlying symmetries. For
example, the backward compatibility of an API can be viewed as a form
of time symmetry: the platform's essential behavior, specifically how
it interacts with older clients, remains invariant despite the passage
of time and internal evolution. In this analogy, the ``conserved
quantity´´ is the client's continued ability to function without
disruption.

\section{Formalizing Symmetry in Software Platforms}
\label{sec:formalizingSymmetryInSoftwarePlatforms}

To analyze software platforms as symmetry-enforcing systems, we define
a \textbf{symmetry of a software platform} as a transformation that
can be applied to the platform’s configuration, environment, or inputs
which does not alter the platform’s essential behavior or
outcomes. Formally, consider a platform $P$ that provides a set of
services. Let $S$ be the state space of the platform and $O$ the set
of observable outputs (or behaviors). A transformation $T: S \to S$ is
said to be a symmetry of the platform if for every state $s \in S$,
the platform’s observable behavior from state $s$ is indistinguishable
from its behavior from state $T(s)$. In other words, $P(s)$ and
$P(T(s))$ produce the same outcomes in $O$. 

The set of all such symmetry transformations of $P$ (with the
operation of composition) forms a \emph{symmetry group} $G$. If $T_1$
and $T_2$ are symmetries, then their composition $T_1 \circ T_2$ is
also a symmetry. A trivial example is the set of all permutations of
$n$ identical server nodes, which is isomorphic to the symmetric group
$S_n$.

We use this model to describe current system behavior and to frame
symmetry as a design objective, treating its equations as conceptual
signposts rather than as tools for explicit calculation.

In practical terms, we identify several categories of transformations:

\begin{itemize}
  \item \textbf{Structural Symmetry:} Swapping equivalent components e.g., identical server nodes does not affect service correctness.
  \item \textbf{External Symmetry:} Clients using the same protocol receive identical treatment, ensuring fairness across tenants.
  \item \textbf{Interface Symmetry:} A uniform API applies the same verbs to all resources, enabling generic tooling and extensibility.
  \item \textbf{Temporal Symmetry:} Idempotent operations and eventual consistency ensure outcomes are invariant to execution order or timing.
  \item \textbf{Quantitative Symmetry:} Scaling out via stateless replication preserves functional correctness independent of resource count.
  \item \textbf{Semantic Equivalence:} Implementation details (e.g., VM migration) do not affect observable behavior at the platform boundary.
\end{itemize}

\begin{table}[h]
  \small  
  \caption{Examples of symmetry types across three different platform domains}
  \label{tab:symmetry-across-domains}
  \begin{tabular}{p{3cm}p{3.2cm}p{3.2cm}p{3.2cm}}
    \toprule
    \textbf{Symmetry Type} & \textbf{Kubernetes} & \textbf{ERP Systems} & \textbf{Social Media Platforms} \\
    \midrule
    Structural & Pod replicas managed by ReplicaSets & Reusable workflow roles (e.g., approver/requester) & Uniform processing stages for posts, comments, and messages \\
    
    Interface  & RESTful API with standard verbs (GET, POST, etc.) & Consistent UI patterns across modules & Unified content ingestion API for varied media types \\
    Temporal  & Idempotent control loops and declarative state reconciliation & Retryable financial operations or form submissions & Time-invariant reprocessing pipelines (e.g., moderation, ranking) \\
    Quantitative & Horizontal scaling of stateless services & Load balancing across functionally identical instances & Sharded processing using uniform microservices \\
    \makecell[l]{Semantic\\Equivalence} & Migration of workloads between nodes without functional change & Interchangeable reporting modules with identical semantics & Alternate ranking models yielding equivalent ordering behaviors \\
        \makecell[l]{Controlled\\ Symmetry\\ Breaking} & Taints and affinity rules for special-purpose nodes & Role hierarchies with elevated privileges & Verified accounts or moderation roles with elevated powers \\
    \bottomrule
  \end{tabular}
\end{table}

The exact form symmetries take will depend on the individual
platform. In table \ref{tab:symmetry-across-domains} we illustrate
different types of symmetries manifest in Kubernetes, ERP systems and
social media platforms.

Symmetry in platforms is closely tied to the notion of
\textbf{conserved properties}. By analogy to Noether’s
theorem\cite{KosmannSchwarzbach2010}, if a platform has a symmetry,
there is some property that remains conserved. If a platform is
invariant under replacing one component implementation with another,
then what is conserved is the \emph{functionality or output} as seen
through the interface. Identifying these invariants tells architects
what aspects of the system can change freely.

It is also important to recognize where symmetry must be
\textbf{broken}. Any non-trivial platform must handle concerns that
are not symmetric. A classic example is security: platforms
break symmetry between an authenticated user and an unauthenticated
one. Performance optimization often leads to asymmetries:
caching some data locally breaks the symmetry between requests that
hit the cache and those that do not, though it conserves overall
correctness. The art of platform design is deciding which symmetries
to enforce globally and where to allow or introduce
asymmetry.

\begin{table}[h]
  \small  
  \caption{Symmetry and related architectural principles}
  \label{tab:symmetry-vs-others}
  \begin{tabular}{p{2.5cm}p{4.5cm}p{6.0cm}}
    \toprule
    \textbf{Concept} & \textbf{Core Focus} & \textbf{How Symmetry Relates or Extends It} \\
    \midrule
    Modularity & Decomposition into components & Asks whether modules are interchangeable under some transformation \\
    Abstraction & Hiding implementation details & Analyzes how abstractions remain valid across system states or roles \\
    Loose Coupling & Minimal interdependency & Focuses on whether loosely coupled parts behave identically when swapped \\
    DRY & Avoiding repetition of knowledge & Investigates when duplication is not only avoidable but collapsible under transformation \\
    \makecell[l]{Substitutability\\(e.g. LSP)} & Replaceability of instances within a type hierarchy & Symmetry generalizes replaceability to whole system components or configurations \\
    \bottomrule
  \end{tabular}
\end{table}

While symmetry intersects with well-established principles such as
modularity, abstraction, and substitutability, it also generalizes
them: modularity decomposes a system, but symmetry asks which parts
are interchangeable; abstraction hides implementation details, while
symmetry examines whether such abstractions remain invariant under
change; substitutability enables replacement within type hierarchies,
whereas symmetry extends this to transformations across
configurations, roles, or time. Viewed this way, symmetry is not
merely adjacent to these principles it may serve as a unifying
abstraction that explains when and why they succeed, and where they
fall short.

\section{Case Study: Kubernetes as a Symmetry-Enforcing Platform}
\label{sec:caseStudyKubernetes}

Kubernetes \cite{Burns2016,KubernetesDocs2023}  is a cloud platform for orchestrating containerized
applications, and it provides a rich example of symmetry enforcement
in practice. It was designed to manage
clusters of machines by abstracting them into a unified compute
substrate, hiding the complexity of
individual machines behind a symmetric, homogeneous facade.

\subsection{API Surface and Resource Model}
Kubernetes presents a \textbf{uniform, resource-based API} to users
and clients. All fundamental entities in Kubernetes (Pods, Services,
Deployments, Nodes, etc.) are represented as \emph{objects} that can
be manipulated through a RESTful interface using standard verbs:
clients can \texttt{GET}, \texttt{POST}, \texttt{PUT}/\texttt{PATCH},
and \texttt{DELETE} these resources. This is a clear instance of
\emph{interface symmetry}. The platform does not require a custom
protocol for each resource type; instead, it has a consistent,
invariant approach. The benefits are evident: tools like
\texttt{kubectl} can be generic. A developer who learns how to manage
a Deployment can apply the same operations to a
PersistentVolume. Kubernetes’ API conventions (the use of \verb|kind|,
\verb|metadata|, \verb|spec|, and \verb|status| fields in nearly all
objects) further reinforce this uniformity and predictability \cite{KubernetesDocs2023}.

The API is extensible in a symmetric way via \textbf{Custom Resource
  Definitions (CRDs)} \cite{KubernetesCRDs2023}. A CRD lets platform operators introduce a new
resource type that behaves like a native Kubernetes object. Once
registered, users can use the same API verbs and tools to manage
it. This preserves the symmetry of the platform’s interface even as
functionality grows, acting as a governance mechanism that channels
extensions through a uniform model.

Furthermore, Kubernetes objects are intended to be \textbf{manipulated
  declaratively}: users provide the desired state in the \verb|spec|
field, and the system works toward making the actual state (reported
in the \verb|status| field) match it. This declarative approach has a
\emph{time symmetry}: whether a configuration is applied in one step
or gradually, the end result is intended to be the same. For example,
creating a Deployment with 3 replicas and later scaling it to 5 yields
the same final state as directly creating it with 5 replicas; the
intermediate path does not matter to the final outcome. The system is
invariant to certain sequences of operations, and the conserved
property is the eventual consistency of the cluster’s state with the
user’s specification.

\subsection{Symmetry in Controller Behavior (Control Loops)}
One of Kubernetes' fundamental design principles is the use of
\textbf{control loops} (controllers) for automating management
tasks \cite{KubernetesControllers2023}. Each controller is a non-terminating loop that observes the
current state of the system and issues actions to drive it toward a
desired state. Notably, this architecture applies uniformly to a wide
variety of tasks: deployment rollout, replica management, and node
health monitoring are all handled by separate controllers that share
the same basic pattern. This is an example of a \emph{structural
symmetry} in the platform’s control logic. Instead of having one-off
scripts, Kubernetes implements each as a controller following the same
design template: watch a resource, and if the actual state diverges
from the declared \verb|spec|, perform operations to reconcile it. All
controllers subscribe to events and update state via the uniform API
server.

Because all controllers adhere to this common pattern, the system
benefits from a form of symmetry at an architectural level. The
mechanism of reconciliation is the same. This uniformity means that
the platform’s behavior (eventual convergence to declared state) is
conserved regardless of which aspect of the system we consider. It
also means that new controllers can be written
independently. Furthermore, there is symmetry between different
instances of the same controller type. The Deployment controller
treats each Deployment object with the same algorithm, ensuring
fairness and predictability.

Temporal symmetry is also evident in
controller operation. Since controllers are constantly watching and
reconciling, the exact timing of an event is not critical if a pod
fails at time $t$ or $t+\Delta$, the replication controller will
notice and restore it. The control loop’s
logic is typically idempotent and retry-driven, providing invariance
to intermittent failures or delays.

\subsection{Scheduling and Node Interchangeability}

The Kubernetes scheduler treats worker nodes as a fungible pool: any Pod may run on any node with sufficient resources, embodying structural symmetry and improving utilization and resilience \cite{KubernetesScheduling2023}. Symmetry can be deliberately broken via \emph{node selectors}, \emph{affinity/anti‑affinity}, and \emph{taints/tolerations} to leverage specialised hardware or enforce policy. Thus interchangeability is the default, and asymmetry is a controlled, explicit choice.

\subsection{Symmetry and Policy in Kubernetes Governance}

Project governance further reinforces symmetry: all API types undergo uniform review, and backward compatibility preserves temporal symmetry by maintaining client functionality across versions \cite{KubernetesGovernance2023}. Operators may still introduce asymmetry e.g., tenant isolation or node pools but these variations are expressed through the same symmetric mechanisms, preventing architectural fragmentation while allowing policy diversity.

\section{Discussion: Implications of Symmetry for Platform Design and Evolution}
\label{sec:discussion}
The case study of Kubernetes, along with the theoretical perspective
developed earlier, suggests that symmetry is a powerful organizing
principle in platform architecture. We now discuss the
broader implications.

\subsection{Benefits of Symmetry in Platform Design}

\textbf{Simplicity and Regularity:} Symmetry imposes a regular
structure on the platform, which can significantly reduce cognitive
load for both designers and users. When a platform behaves uniformly,
one can predict its behavior in new situations by analogy. This is the
DRY (Don’t Repeat Yourself) principle at the system level. For
instance, the uniform API of Kubernetes means that a single library
can be written to interact
with any part of the system. The regularity provided by symmetry often
yields a \emph{principle of least surprise}.

\textbf{Reusability and Composability:} Symmetric interfaces and
behaviors allow components to be reused and composed more easily. If
two components adhere to the same interface, a third component can
interact with either one without special adaptation. This enables
ecosystems: third-party tools can integrate with the platform if it
consistently follows certain conventions. The Cloud Native ecosystem
around Kubernetes thrives because tools like monitoring agents and
GitOps controllers all integrate by reading or writing symmetric API
objects.

\textbf{Scalability and Fault Tolerance:} A symmetric architecture
facilitates horizontal scaling. If every node is running identical
software, scaling out is simply a matter of adding more nodes. There
is no single master bottleneck (unless the master is a separate
role). Similarly, fault tolerance is improved when roles are
symmetric: the failure of any one component is handled by another
equivalent component. In contrast, if the system relies on a unique
component (breaking symmetry), that component’s failure is more
disruptive. Symmetrical redundancy is a key strategy in reliable
systems.

\textbf{Ease of evolution:} Platforms inevitably evolve. Symmetry can
make evolution smoother by localizing changes. If the platform has a
well-defined invariant interface, one can improve internals without
affecting users. This treats an internal change as a symmetry
transformation that leaves external behavior invariant. Moreover,
adding new features in a symmetric way (e.g., via CRDs) means the
platform can grow without losing coherence. Symmetry provides a stable
“core” that can be extended, a concept echoed in product platform
theory.

\subsection{The Need for Symmetry Breaking}

While symmetry provides many benefits, a perfectly symmetric system is
often suboptimal. In practice, certain requirements demand breaking
symmetry.

\textbf{Performance optimizations and locality:} Real-world systems
are not uniform. To achieve optimal performance, platforms might break
symmetry by introducing special-case handling. For example, a database
might cache hot rows (treating them differently from cold data), or a
distributed storage system might keep an extra replica in a nearby
location. These optimizations create asymmetries but are necessary for
efficiency. The key is that when breaking symmetries, this should
ideally be managed internally, and not violate the platform’s promises
from an external perspective.

\textbf{Security and control:} As
mentioned, not all actors can be treated equally. Roles and permissions
intentionally break the symmetry among users. Multi-tenant platforms often partition
resources so that tenants are isolated; this breaks the symmetry of
the resource pool. These measures are
crucial for safety and fairness. While global symmetry is broken, each tenant may see a symmetric platform within their own scope.

\textbf{Special hardware and capabilities:} Modern infrastructures
include specialized hardware like GPUs or FPGAs. Utilizing these
efficiently means breaking the assumption that every node is
identical. Platforms need ways to schedule work to specific
hardware. Kubernetes’ taints and node selectors are a case in
point. They break the homogeneous cluster symmetry to allow
specialized scheduling, but the platform retains as much API symmetry
as possible.

The act of symmetry breaking in design should be done
\textbf{intentionally and minimally}. A useful guiding principle is to
first build the system under the strongest reasonable symmetry
assumptions, then introduce asymmetry only to address concrete
issues. This echoes how patterns are applied to break the ''land
uniformity´´ of a naive design just enough to handle a requirement.

\subsection{Governance and Evolution}

In this framework, governance acts as a symmetry-preserving mechanism:
it enforces architectural invariants, prevents uncontrolled
asymmetries, and guides extension mechanisms to remain coherent with
the platform’s underlying symmetry assumptions.  If too many ad-hoc
changes are introduced, the platform risks losing its elegant
invariants and becoming a patchwork of exceptions. Good governance
will enforce that contributions align with the platform’s core design
principles, many of which are symmetry-related.

At the same time, governance must also manage the introduction of
necessary asymmetries. This includes communicating changes, ensuring
backward compatibility, and carefully designing extension points. One
might draw an analogy to constitutional law: a platform’s architecture
sets out the “constitution” of its symmetric assumptions. Governance
ensures that new laws (features) don’t violate the constitution
without due process.

A concrete example is the decision in Kubernetes to move from
PodSecurityPolicies (PSP) to Pod Security Standards. PSP was a
mechanism that somewhat inconsistently bolted on security rules. The
community decided to replace it with a more uniform mechanism that
uses namespaces and labels concepts that are already symmetric and
pervasive in Kubernetes. This is a case of restoring symmetry by
removing a feature that was conceptually asymmetric. The platform’s
evolution in this case favored symmetry and consistency.

\section{Conclusion and Future Work}
\label{sec:conclusionsAndFutureWork}

In this paper, we examined software platforms through the lens of
symmetry, arguing that many successful platforms function as
symmetry-enforcing systems. We surveyed literature, formalized the
concept of symmetry as invariance under transformation, and used
Kubernetes as a case study. We demonstrated how a real-world
platform enforces symmetry: its API is uniform, its nodes and
controllers are treated uniformly, and its design encourages
interchangeable components. We also observed where Kubernetes
intentionally breaks symmetry (e.g., distinguishing control plane and
worker roles, or using scheduling constraints) to meet practical
demands, reflecting the nuanced balance required in robust
architectures.

The implications of treating symmetry as a first-class design
principle are significant. Symmetry contributes to modularity,
scalability, and ease of evolution. It can serve as a guide during
design and a rationale for resisting ad-hoc changes in platform
governance. However, we also highlighted the importance of intentional
symmetry breaking for handling specialization and optimization.

There are several avenues for future research building on this work.

One direction is to develop \textbf{metrics or formalisms for symmetry
  in software architectures}. Just as one can quantify complexity,
perhaps one can quantify symmetry (or deviation from it) in a
design. For example, graph-based models of an architecture could be
analyzed for automorphisms (symmetry mappings) to detect how
symmetrical a system is. This could potentially inform complexity
measures or identify refactoring opportunities.

Another area is exploring \textbf{symmetry in other platforms and at
  other scales}. While we focused on Kubernetes, the concept could be
applied to high-level platforms like mobile app ecosystems or browser
engines. Do these exhibit symmetry-enforcing behavior (e.g., all apps
in a mobile OS are managed the same way for fairness and security)? A
comparative study could validate and enrich the symmetry hypothesis.

Additionally, investigating \textbf{category theory or advanced
  algebraic approaches} might yield deeper insights. Category theory
could model not just symmetries (automorphisms) but relationships
between different symmetric structures (functors). While we are no
fans of premature formalism, future work might show clear value in
reasoning about families of platforms or cross-platform invariants.

Finally, there is practical work in building \textbf{tooling for symmetry-aware
  design}. For instance, design frameworks or language features that
encourage defining invariants and supporting symmetrical extension
could help architects. In configuration management, one might create
tools to automatically ensure consistency across
environments, or in microservice design, tools to enforce that all
services expose a common observability interface as a symmetric
concern.

In conclusion, viewing software platforms as symmetry-enforcing
systems offers a unifying perspective that connects architectural
principles with deep concepts from mathematics. By consciously
applying symmetry where appropriate and symmetry breaking where
necessary, platform designers can better manage complexity and change.

\bibliographystyle{abbrv}
\bibliography{platform-as-symmetry-arxive}

\end{document}